# Index Selection for NoSQL Database with Deep Reinforcement Learning

Shun Yao
School of Computer Science
Harbin Institute of Technology
Harbin, China
shonyao@outlook.com

Hongzhi Wang
School of Computer Science
Harbin Institute of Technology
Harbin, China
wangzh@hit.edu.cn

Yu Yan
School of Computer Science
Harbin Institute of Technology
Harbin, China
yuyan0618@hit.edu.cn

**ABSTRACT**

We propose a new approach of NoSQL database index selection. For different workloads, we select different indexes and their different parameters to optimize the database performance. The approach builds a deep reinforcement learning model to select an optimal index for a given fixed workload and adapts to a changing workload. Experimental results show that, Deep Reinforcement Learning Index Selection Approach (DRLISA) has improved performance to varying degrees according to traditional single index structures.

## 1 INTRODUCTION

The development of the Internet and cloud computing needs databases to be able to store and process big data effectively. With the demand for high-performance when reading and writing, NoSQL databases are more and more widely used. So far, more than 225 different NoSQL stores have been reported and the list is still growing [1].

In NoSQL databases, it is essential to provide scalable and efficient index services for real-time data analysis. The index selection plays an important role in NoSQL database performance improvement. Only if we can correctly select the index that fits the workload, the performance of the database is effectively guaranteed.

Index selection needs to be compatible with most workloads and as many indexes as possible. In most of the time, the workload is not immutable, but will change while database is being used. Besides, the optimal index selections under different hardware environments are have wide differences.

Machine learning has made great strides in recent years, which provide new opportunities for index selection. Most existing index solutions focus on improving write or read throughput. And many researchers propose various new index structures to improve database performance [2-8]. However, the current research work on the index selection usually use one singe immutable index configuration. However, single index structure can hardly deal with workloads in many different situations. Additionally, there is no existing NoSQL database solution combining several different index structures at the same time, which is common in practice.

Different from existing methods, we focus on a general index selection method which fits various workloads under different hardware environments. In this paper, we explore selecting optimal index structure and tuning its parameters dynamically at the same time by using reinforcement learning. There are still some potential problems in the process of using deep reinforcement learning model for the intelligent selection decision of index structure. First, defining the environment required for deep reinforcement learning in our situation is non-trivial. The reason is that there is no intuitive relationship between index selection and the state and actions of reinforcement learning. In addition, how to set the action of reinforcement learning and its corresponding reward function is also very worth exploring, since a bad cost function will make the reinforcement learning model difficult to converge. After that, we should get optimal index structure and its' parameter from Q value learned in the deep reinforcement learning model. And it is worth noting that the local optimal solution that selects the maximum Q value action every time may not be the global optimal solution.

Our solution to NoSQL index selection problem and our developed Deep Reinforcement Learning Index Selection Approach (DRLISA) are distinct from the related work in several ways. The uniqueness of our approach relies in combination of:

- Recommendation of index parameters for proposed indexes while recommending the index structure.
- Using reinforcement learning model as a heuristic method to deal with index selection under the dynamic workload.
- Selecting index based on Q value from reinforcement learning network.

In existing modern databases, the means for self-tuning specialized indexes are provided. Nevertheless, none of the present index selection approaches has recommended a general method of the recommendation of index parameters. They only recommend index parameters of one specialized index. Whereas in this paper, we first introduce a general reinforcement learning method to select the index and its parameters and then in our approach, we deal with the situation with dynamic workload. Having such features in such a database management accomplice will not only reduce the overhead for the administrator of the system to choose proper indexes, but also provide a good skeleton for developing an automatic index selection system.

In the following sections, we will discuss the specific implementation of our index selection method. Section 2 provides the necessary background of our index selection architecture. Section 3 shows our design of the index selection architecture in details. Section 4 is about the testing of our selected index, showing how it outperformed than singe immutable indexes.

## 2 BACKGROUND

**Reinforcement Learning.** Reinforcement learning models are able to map scenarios to appropriate actions, with the goal of maximizing a cumulative reward [9]. At each timestep, $t$, the agent will observe a state of the environment, $s_t$ and will select an action, $a_t$. The action selected depends on the policy, $\pi$. This policy can reenact several types of behaviors. As an example, it can either act greedily or balance between exploration and exploitation through an $\epsilon$-greedy (or better) approach. The policy is driven by the expected rewards of each state, which the model must learn. Given the action selected, the model will arrive at a new state, $s_{t+1}$. The environment then sends the agent a reward, $r_{t+1}$, signaling the "goodness" of the action selected. The agent's goal is to maximize this total reward [9]. One approach is to use a value-based iteration technique, where the model records state-action values, $QL(s, a)$. These values specify the long-term desirability of the state by taking into account the rewards for the states that are likely to follow [9].

**Deep Q Networks.** A deep Q network (DQN) is a multi-layered neural network that for a given state $s$ outputs a vector of action values $Q(s, a; \theta)$, where $\theta$ are the parameters of the network [10]. Two important ingredients of the DQN algorithm as proposed by Mnih et al. (2015) are the use of a target network, and the use of experience replay [11]. And both the target network and the experience replay dramatically improve the performance of the algorithm.

**Dueling Network.** The different between dueling network and DQN is dueling network split one estimator into two separate estimators: one for the state value function and one for the state-dependent action advantage function [12]. The main benefit of this factoring is to generalize learning across actions without imposing any change to the underlying reinforcement learning algorithm. The architecture leads to better policy evaluation in the presence of many similar-valued actions.

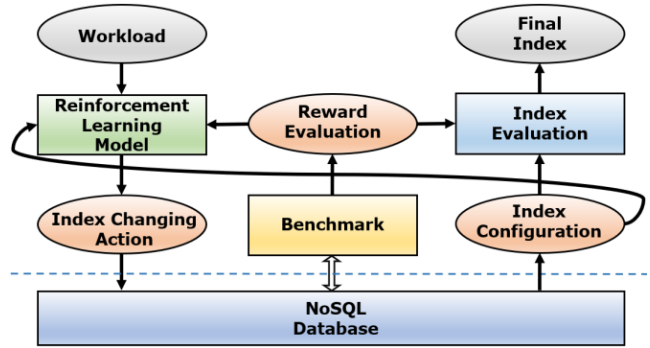

**Figure 1: The Architecture of DRLISA**

## 3 NOSQL DEEP REINFORCEMENT LEARNING INDEX SELECTION APPROACH (DRLISA)

In this paper, we use the term configuration to mean an index and its parameters. The goal of a DRLISA is to select an index configuration that is as close to optimal index configuration as possible for given database and workload, where workload consists of a set of NoSQL data manipulation statements. Section 3.1 overviews DRLISA. The details of Reward Evaluation module are shown in Section 3.2. Section 3.3 introduces the reinforcement learning model. In Section 3.4, we discuss how to select the optimal index after training model. Besides, Section 3.5 is the discussion about DRLISA.

### 3.1 Overview

An overview of the architecture of DRLISA is shown in Figure 1. The dotted line in the figure denotes the process boundary between our modules and *NoSQL Database*. The index selection tool takes as input a workload on a specified *NoSQL database*.

This approach has a basic *Reinforcement Learning Model* module. The *Reinforcement Learning Model* module takes as input workload and index configuration representations, and it recommends an action from the action set. The reward and the model will be introduced in Section 3.2 and 3.3, respectively.

Revolving around the *Reinforcement Learning Model* module, other modules of DRLISA are as follows.

The *Index Changing Action* module is responsible for taking actions from the *Reinforcement Learning Model* module and executing them in the *NoSQL Database*.

We use *Benchmark* module to get the index performance from *NoSQL Database*. And *Reward Evaluation* module calculates the reward value of an index changing action by using index performance we got from *Benchmark* module.

The *Index Configuration* module get the index configuration representation of current index structure and related parameters from NoSQL database. *Index Evaluation* module is used for selecting the final optimal index configuration according to the reward from *Cost Evaluation* module and the index configuration representation from *Index Configuration* module.

The details of *Reward Evaluation* module, *Reinforcement Learning Model* module, and *Index Configuration* module will be introduced in section 3.2, 3.3 and 3.4, respectively.

## 3.2 Reward Evaluation

Reward function is the most important part of each reinforcement learning model. The reasonableness of reward function generally determines the reinforcement learning model is converge or not. The reward function should give the value of actions based on the information fed back by the environment. Thus, in this section, we focus on the reward evaluation.

The reward function $r_t$ of an index changing action of time $t$ should consider time reward $r_{time}(t)$, and switch index cost $c_{switch}(t)$ of a new index. And the performance, which we can get from the *Benchmark* module by testing data under the given workload, would help us to calculate these functions above. If the action keeps the index configuration, the reward function is set to zero. $s_t$ is the state representation of time $t$ including index configuration and workload, and the reward function of an action of time t is defined as

$$r_t = \begin{cases} r_{time}(t) - c_{switch}(t), & s_t \neq s_{t-1} \\ 0, & s_t = s_{t-1} \end{cases}$$

The time reward function $r_{time}(t)$ needs to show the difference between the time cost of time $t-1$ and time $t$. Compared with the method of directly calculating the function with time cost, the calculation with throughput can better reduce the error and is not affected by the variable of the number of operations. According to the throughput $p_t$ of time t representing the number of NoSQL database operations per second, which we can get from *Benchmark* module, where $k$ is a constant and the time reward function $r_{time}(t)$ of time $t$ is defined as follow

$$r_{time}(t) = \begin{cases} ln\left(\frac{(p_t - p_{(t-1)})}{k} + 1\right), & p_t > p_{t-1} \\ 0, & p_t \leq p_{t-1} \end{cases}$$

The index switching cost $c_{switch}(t)$ should be a small negative value compared with time reward function $r_{time}(t)$. By adding a small negative index switching cost we can reduce the actions that make time reward function larger. Besides, we can also add space cost function into reward function when we need to consider the space cost.

## 3.3 Reinforcement Learning Model

The reinforcement learning model module takes as input state representation $s = \{N, C\}$, where $N$ and $C$ represent the workload representation and index configuration representation, respectively. The output of reinforcement learning model module is Q values of each actions from action set. The action set includes several different actions like keeping the current configuration, changing the index structure, and changing the index parameters.

For index selection, the value of state representation is more important than action advantage. In our index selection tool, we use the dueling network architecture as the neuron network architecture of reinforcement learning model to train faster and make the model performance better. The reason is that, distinct from DQN, dueling network architecture has an independent fully connected layers to output a scalar representing state value, which fits our problem well. With such independent layers, we can get more accurate state values.

Because of the positive value of the reward function we used, the learning of the reinforcement learning model might show the phenomenon of cyclical scoring. The actions it used will be cyclical, and the discounted return $R_t$ also increases. Obviously, we do not want to see this phenomenon.

In each episode, we store the appeared state list $L$ to recording all appeared state, and it could avoid the cyclical actions. When one of configuration representations appears twice, we break the episode. The reason why it works is that the derivative of the cost function $r_t$ we used near zero is larger than away from zero, and the minus between throughput $p_t$ is linear. The optimal strategy to get maximum discounted return $R_t$ is making the index better gradually. Then we can get optimal index configuration by finding maximum discounted return $R_t$.

---

**Algorithm 1** Training the Deep Reinforcement Learning Model

**Inputs:** Workload $N$
1: Initialize the environment $Env$
2: Build deep reinforcement learning neuron network architecture $RL$
3: **for** $episode = 0 \dots num - 1$ **do**
4:    Reset the environment $Env$ with workload $N$ and a random initial index configuration $C$
5:    Get the current state $s$ from environment
6:    Clear the appeared state list $L$
7:    **repeat**
8:      Add current state $s$ to appeared state list $L$
9:      Choose action by run $RL$ with current state $s$
10:     Get next state $s'$ and reward from $Env$
11:     Store transition $(s, a, r, s')$ into experience pool
12:     Sampling transitions to update network each several steps
13:     $s = s'$
14:    **until** current state $s$ is already in the appeared state list $L$

---

We propose Algorithm 1 to train deep reinforcement learning models. We need an initial workload as input to the algorithm. At the beginning, Line 1-2 initialize the environment required for reinforcement learning and the deep reinforcement learning neural network. After that Line 3-14 perform many episodes.

For every episode, Line 4 resets the workload and a new random index configuration in the environment. Then we extract the current state from the environment and clear the list of states appeared in Line 5-6. Line 7-14 perform many steps in each episode.

At the beginning of each step, Line 8 adds the current state to the list of states appeared. After passing the current state to the neural network, we get recommended actions in Line 9. Performing actions in the environment, we can get the next state and the reward value of this action in Line 10. Since we need to train the neural network model, Line 11 puts the current state, the next state, this action and the reward value of this action as a transition into the experience pool. After several steps, Line 12 extracts some transitions from the experience pool for neural network learning. Line 13 assign next state to current state to complete one step. Line 14 repeats each step in Line 8-13 until the current state appears in the list of states appeared.

After executing all episodes, the reinforcement learning neural network training corresponding to the workload is completed.

The time complexity of the algorithm is $O(P * (T_a + T_{nn}))$, where $P$ is the total number of training steps, $T_a$ is the time complexity required to perform actions in the environment, and $T_{nn}$ is affected by different neural network structures, including the time required for the neural network to propagate forward and backward.

### 3.4 Index Evaluation

After the training of the reinforcement learning model completed, we only have the value corresponding to each action. And what we need is the optimal index configuration. The *Index Evaluation* module gives the approach to selecting the optimal index configuration as follows.

The algorithm selecting the optimal index configuration has a little different from the algorithm in Training the Deep Reinforcement Learning Model. In each episode, we discount the reward of each action, and we record the current state and its corresponding discounted return. The discounted return can help us to find the final optimal index configuration from the reinforcement learning model. Initializing a random configuration, we can choose action by using the reinforcement learning model until an episode end. Repeat this process and we could get some configurations and their corresponding discounted returns. What we need to do is select the configuration which have maximum discounted return.

Algorithm 2 selects the final index configuration based on the deep reinforcement learning model. The input to the algorithm is a workload. Initialization in Line 1-3 is similar to the initialization of the training reinforcement learning model, the difference is that we need to additionally initialize the optimal index configuration and its corresponding maximum discounted reward in Line 3. After that Line 4-18 perform many episodes.

**Algorithm 2** Selecting the Optimal Index Configuration
**Inputs:** Workload $N$
1: Initialize the environment $Env$
2: Load deep reinforcement learning neuron network $RL$
3: Initialize the optimal index configuration $C_{optimal}$ and optimal maximum discounted return $R_{max}$
4: **for** $episode = 0 \ldots num - 1$ **do**
5:    Reset the environment $Env$ with workload $N$ and a random initial index configuration $C$
6:    Get the current state $s$ from environment $Env$
7:    Clear the appeared state list $L$
8:    Reset current discounted return $R$
9:    **repeat**
10:      **if** $R_{max} < R$ **then**
11:        $R_{max} = R$
12:        $C_{optimal} = C$
13:      Add current state $s$ to appeared state list $L$
14:      Choose *action* by run $RL$ with current state $s$
15:      Get next state $s'$ and reward $r$ from $Env$
16:      $R = R + r$
17:      $s = s'$
18:    **until** current state $s$ is already in the *appeared* state list $L$
19: **return** $C_{optimal}$

The initialization of each episode is also the same as the initialization of the training reinforcement learning model in Line 5-7. We need to additionally initialize the current discounted reward to zero in Line 8. Line 9-18 performs many steps in each episode.

At the beginning of each step, Line 10-12 update the maximum discounted reward of the optimal index configuration with the current discounted reward, while maintaining the optimal index configuration. Then Line 13 adds the current state to the list of states appeared. After passing the current state to the neural network, we can obtain recommended actions in Line 14. With the actions in the environment, we can get the next state and the reward value of this action in Line 15. Line 16 adds the current discounted reward to the reward value of this action. Line 17 assign next state to current state to complete one step. Line 18 repeat each step until the current state appears in the list of states that have occurred.

After executing all episodes, the final optimal index configuration is returned as the algorithm output in Line 19.

The time complexity of the algorithm is $O(P' * (T_a + T_{nn}))$, where $P'$ is the total number of trying steps, $T_a$ is the time complexity required to perform actions in the environment, and $T_{nn}$ is affected by different neural network structures, including the time required for the neural network to propagate forward and backward.

## 3.5 Discussion

DRLISA select the index configuration more reasonable than selecting index manually, and it could deal with the NoSQL database intelligent index selection problem. Different from single index structure, DRLISA supports selecting index structure and its parameter under dynamic workloads. Besides, DRLISA shows its strong scalability, reflected in that we can add other indexes into the index set, add actions into action set or modify reward evaluation function in different situations as extensions. The strong scalability DRLISA is consistent with the concept of NoSQL database. We believe that Deep Reinforcement Learning Index Selection Approach (DRLISA) can bring new inspiration to NoSQL index selection problem.

## 4 EXPERIMENT

We now show the practical performance of DRLISA. Section 4.1 gives a brief introduction to each part of NoSQL DRLISA architecture. Section 4.2 presents the details of training model. Experiment results are presented in section 4.3.

### 4.1 Experimental Setup

The benchmark is YCSB (Yahoo! Cloud Serving Benchmark) [13], whose workload contains common NoSQL data manipulation statements, i.e., Insert, Update, Scan, Read, RMW(Read-Modify-Write). In our experiments, we use Python TensorFlow to implement our approach [14].

The database used is PostgreSQL with a Foreign Data Wrapper for RocksDB. We test common-used index structures including B-Tree, Hash and LSM-Tree.

We have experimented this proposed model on a system having 8 processors with 8 GB RAM. This server is running with Ubuntu 18.04 LTS.

### 4.2 Training Model

In our experiment, the learning rate is initially set to 0.001, the reward decay is set to 0.7, epsilon-greedy is set to 0.7 and the memory size is set to 50000. And the number of neurons in three hidden layers are 16, 8 and 8, respectively. Initially, we train five workloads, which have only one data manipulation, with 1000 episodes each. And then we select 150 random workloads with 300 episodes to make the model more universal.

### 4.3 Experiment Results and Discussion

**The Impact of Data Manipulation.** Under single data manipulation workload trained, performance of DRLISA and three single traditional index structures are showed in Table 1. The index selected by our approach has better performance than a single index including B-Tree, Hash and LSM-Tree. We can see that single index structures unavoidably have shortcomings in certain data manipulation. And DRLISA can merge their strengths.

| Workloads | Throughput of indexes | | | |
|---|---|---|---|---|
| | Selected | B-Tree | Hash | LSM-Tree |
| Read | 2702.7 | 2762.4 | 2688.2 | 21.9 |
| Update | 35.4 | 39.2 | 38.8 | 18.0 |
| Scan | 4524.9 | 4048.6 | 4065.0 | 30.1 |
| Insert | 2415.5 | 39.0 | 39.3 | 2352.9 |
| RMW | 38.7 | 36.3 | 38.3 | 10.1 |

Table 1: The Impact of Data Manipulation.

**The Impact of Operation Count.** Figure 3 shows impact of count of operations on throughput of the workload that the insertion proportion is 100%. The relatively stable throughput which indicates the stability of the test environment.

**Comparison with Single Index Structure.** For some workloads we never trained before, the approach is also worked and the performance we got is better than single traditional indexes structures. We test 100 workloads, and the throughput by using our method can improve 3.25% compared with B-Tree, and 3.19% compared with Hash, 20.15% compared with LSM-Tree. Table 2 shows performance comparison between index selected by DRLISA and single traditional index structures under several workloads.

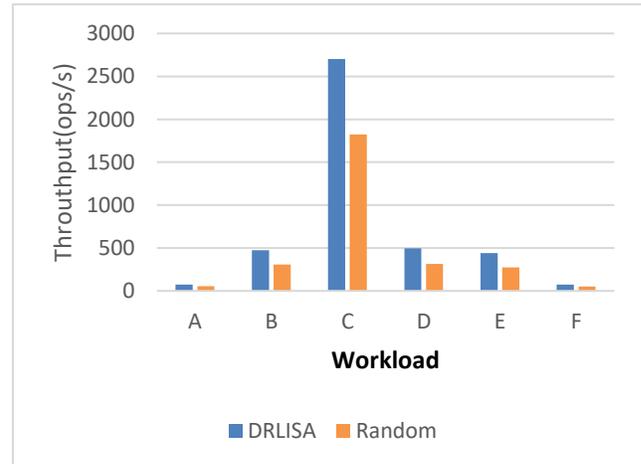

Figure 2: Comparison with Random Selection.

**Comparison with Random Selection.** For the common NoSQL workload, in Figure 2, we test the performance comparison between the indexes selected by DRLISA and randomly selected indexes under the six workloads provided by YCSB by default. Results shows DRLISA is effective.

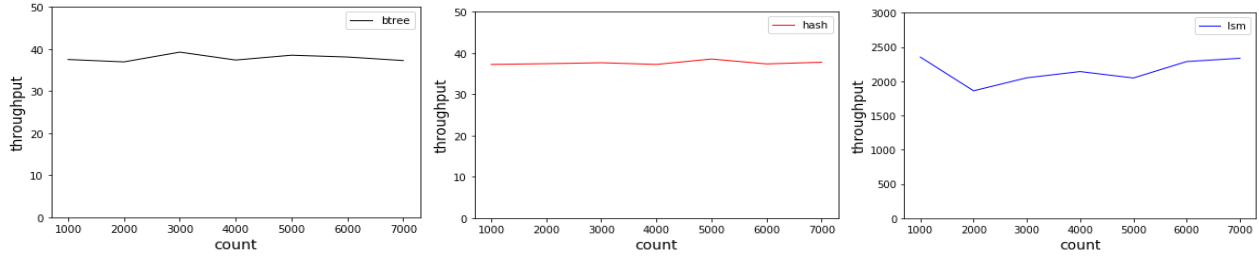

**Figure 3: The Impact of Operation Count.**

| Workloads | | | | | Throughput of indexes | | | |
|---|---|---|---|---|---|---|---|---|
| Read | Update | Scan | Insert | RMW | Selected | B-Tree | Hash | LSM |
| 50% | 50% | 0% | 0% | 0% | 74.25007 | 71.53076 | 71.23014 | 18.51235 |
| 95% | 0% | 0% | 5% | 0% | 496.0317 | 463.6069 | 460.6172 | 18.99011 |
| 0% | 0% | 95% | 5% | 0% | 442.2822 | 379.0751 | 419.8153 | 18.16168 |
| 50% | 0% | 0% | 0% | 50% | 72.64274 | 71.33176 | 71.67431 | 12.11504 |
| 7% | 22% | 18% | 31% | 22% | 51.02301 | 48.00077 | 50.29422 | 17.21911 |
| 28% | 51% | 16% | 1% | 4% | 67.80121 | 63.46386 | 66.04147 | 17.55187 |
| 36% | 47% | 2% | 9% | 6% | 61.32712 | 59.60897 | 58.93099 | 17.74528 |

**Table 2: Comparison with Single Index Structure.**

**Summary.** Through experiments, we confirm that NoSQL deep reinforcement learning index selection method (DRLISA) has good performance under different workload. After sufficient training, under the changing workload, the performance of NoSQL deep reinforcement learning index selection method (DRLISA) is improved to different degrees according to the traditional single index structure. For the common NoSQL database workload, NoSQL deep reinforcement learning index selection method (DRLISA) has a good performance. Therefore, the index selection method combined with deep reinforcement learning can effectively select the index corresponding to the workload.

## 5  CONCLUSION

In this work, we described a model that uses deep reinforcement learning for index selection for NoSQL database. Based on our index selection architecture, we use deep Q network with dueling network architecture to incrementally learn actions of workloads.

As future work, we propose to use the index selection architecture that with reinforcement learning in conjunction with traditional index structure and other new index structures like learned index structure, etc. to learn optimal index for SQL or NoSQL databases.